\documentclass[12pt, preprint]{aastex62}

\newcommand{\mass}{\,$M_{500}$}	
\newcommand{\lum}{\,$L_{X, 500}$}	
\newcommand{\rad}{\,$R_{500}$}  
\newcommand{\psec}{\,$s^{-1}$}   
\newcommand{\tenf}{\,$10^{44}$}  
\newcommand{\Mr}{\,$M_r$}   

\newcommand{\msun}{\hbox{${M}_{\odot}$}}

\newcommand{\be}{\begin{equation}}
\newcommand{\ee}{\end{equation}}
\newcommand{\ba}{\begin{eqnarray}}
\newcommand{\ea}{\end{eqnarray}}

\newcommand{\simgt}{\lower 2pt \hbox{$\, \buildrel {\scriptstyle >}\over {\scriptstyle\sim}\,$}}
\newcommand{\simlt}{\lower 2pt \hbox{$\, \buildrel {\scriptstyle <}\over {\scriptstyle\sim}\,$}}
\newcommand{\ls}{\lower 2pt \hbox{$\;\scriptscriptstyle \buildrel<\over\sim\;$}}
\newcommand{\gs}{\lower 2pt \hbox{$\;\scriptscriptstyle \buildrel>\over\sim\;$}}

\newcommand{\comment}[1]{}  

\graphicspath{{./}{figures/}}

\submitjournal{AJ}

\shorttitle{AGN Fraction in Clusters}
\shortauthors{Mishra \& Dai}

\begin{document}

\title{Lower AGN Abundance in Galaxy Clusters at $z<0.5$}

\email{hora.mishra@ou.edu}

\author[0000-0002-0786-7307]{Hora D. Mishra}
\affil{Homer L.\ Dodge Department of Physics and Astronomy,
University of Oklahoma, Norman, OK 73019, USA}

\author[0000-0001-9203-2808]{Xinyu Dai}
\affil{Homer L.\ Dodge Department of Physics and Astronomy,
University of Oklahoma, Norman, OK 73019, USA}
\email{xdai@ou.edu}



\begin{abstract}

Most galaxies in clusters have supermassive black holes at their center, and a fraction of those supermassive black holes show strong activity. These active galactic nuclei (AGNs) are an important probe of environmental dependence of galaxy evolution, intracluster medium, and cluster-scale feedback. 
We investigated AGN fraction in one of the largest samples of X-ray selected clusters from the \emph{ROSAT} and their immediate surrounding field regions below $z<0.5$. 
We found lower average AGN fraction in clusters (2.37$\pm$0.39)\% than for the fields (5.12$\pm$0.16)\%. 
The lower AGN fractions in clusters were measured, after dividing the clusters into five redshift intervals between 0.0 and 0.5, in each redshift interval, and we found an increase in the fraction for both cluster and field galaxies with redshift below $z<0.5$, which clearly indicates an environment and redshift dependence. 
We further divided the clusters into low-mass and high-mass objects using a mass cut at log$_{10}$(M$_{500}$/\msun) = 13.5, finding comparable AGN fractions for both classifications, while a significantly higher AGN fraction in field. 
We also measured increasing AGN fractions with clustercentric distance for all redshift bins, further confirming the environmental dependence of AGN activities. 
In addition, we did not find an obvious trend between AGN fraction and SDSS \Mr{} absolute magnitudes among different redshift bins.
We conclude that the lower AGN fraction in clusters relative to fields indicate that factors, such as inefficient galaxy mergers and ram pressure stripping cause a deficit of cold gas available in high density regions to fuel the central supermassive black hole. Clusters and fields in present universe have lost more gas relative to their high redshift counterparts resulting in a lower AGN fraction observed today. 

\end{abstract}



\section{Introduction} \label{sec:intro}

Most, if not all, galaxies have a supermassive blackhole (SMBH) at their center. Active galaxies are the ones that host active galactic nulei (AGNs) where the supermassive black hole is accreting mass from the dense central region of the galaxy at a sufficiently high rate. For the luminous sample, the radiation from these supermassive black holes can outshine the entire host galaxy. It is not well understood why only a few SMBHs show such high nuclear activity. Galaxies are found in overdense regions of the universe which form the large-scale structures called galaxy clusters and also in very low density regions of filaments and voids.  There is strong evidence that galaxy evolution is closely related to its environment, influencing star formation, stellar mass, and galaxy morphology (e.g. \citealt{2016ApJ...825...72A}). Most massive early-type galaxies are typically found in galaxy groups and clusters, close to the center of a cluster's gravitational potential. However, large isolated galaxies have also been observed in the universe.  
While it is generally accepted that environment plays an important role in galaxy evolution, its role in triggering nuclear activity in active galaxies is still widely debated. The fraction of galaxies in clusters and fields that host these active SMBHs is an important probe of AGN fueling processes, cold interstellar medium at the centers of galaxies, galaxy's star formation rate, co-evolution of black holes and galaxies, and AGN influence on the local environment via the AGN feedback mechanism (e.g., \citealt{1988ApJ...325...74S}; \citealt{2008ApJS..175..356H}). Powerful jets in radio-loud AGNs and/or AGN outflows interact with the surrounding interstellar medium and can affect star formation rate in the galaxies hosting AGNs. Various local and global environmental factors, such as stellar mass density, distance from the cluster's center, cluster's velocity dispersion, cluster dynamics, host galaxy's morphology, and cosmological redshift may influece AGN fraction in clusters; however the relation between the two is not well understood. Many studies such as \citet{2017MNRAS.472..409L}, \citet{2018A&A...620A.113A}, and \citet{2019MNRAS.487.2491E} find evidence for variation in fraction of AGNs with high and low-density environments, while several others see no to very weak correlation between the two (e.g. \citealt{2003ApJ...597..142M}; \citealt{2013MNRAS.429.1827P}; \citealt{2018MNRAS.475.1887Y}; \citealt{2019MNRAS.tmp.1665M}).    

Several possible triggering mechanisms in clusters and fields to turn on the nuclear activity have been suggested, such as major and minor mergers (e.g. \citealt{2005Natur.433..604D}; \citealt{2015MNRAS.451.2968F}), disc instability (\citealt{2009ApJ...703..785D}), bar influence (\citealt{2012A&A...540A..23S}; \citealt{2019ApJ...877...52Z}), and tidal effects (\citealt{1996Natur.379..613M}). Understanding the triggering mechanisms for these nuclei is crucial to the study of galaxy formation and evolution in the context of its environment. These processes are responsible for supplying gas for accretion onto the black hole, thus triggering it. Due to different accretion rates needed for different luminosity classes, it has been suggested that different physical processes are responsible for triggering the AGN. Of these processes, galaxy interactions and mergers have been thought to be especially important for high luminosity AGNs. Accretion of mass onto the black hole is possible via non-axisymmetric perturbation that triggers the AGN nuclear activity which has been observed to occur during mergers (\citealt{1998ApJ...504..671K}). Numerical simulations have shown tidal torques during mergers to efficiently move the gas inward to fuel rapid black hole mass accretion (e.g. \citealt{1989Natur.340..687H}; \citealt{2005MNRAS.361..776S}; \citealt{2012NewAR..56...93A}). On the other hand, studies have found secular processes, such as minor mergers and disk instability to be the dominant mechanisms for triggering low luminosity AGNs (\citealt{2009ApJ...698.1550H}; \citealt{Hopkins14}).
Since clusters have significantly higher galaxy densities than fields, the rate of mergers is higher in clusters which would seem to trigger more AGNs in clusters. However, a combination of galaxy rich environment, extreme conditions in the cluster's gravitational potential well, star formation rate, and concentration of cold gas in the cluster halo make the AGN dynamics more complex. Pressure from the hot intracluster medium (ICM) may cause suppressed AGN activity since it causes evaporation of the cold gas that accretes in the disk around the SMBH (\citealt{1972ApJ...176....1G}). Studies show star formation to strongly co-evolve with AGN evolution (\citealt{2016ApJ...825...72A}). Cold intracluster gas that drives star formation is also the primary fuel for AGN activity. Influence of galaxy mergers containing rich gas is similar on star formation in those galaxies to that on AGN fraction. Cluster halos, on the other hand, quench AGN formation by capturing cold gas and preventing accretion in the inner parts of the cluster by a process called “strangulation” (\citealt{1980ApJ...237..692L}). In the AGN phase model where most black holes in galaxies undergo an intense activity period, the lifetimes of emission at AGN luminosities are estimated to be in the range $10^6-10^8$ years. Based on models of black hole growth via gas inflows, the strong accretion phase lasts for $\sim 10^8$ years (\citealt{2000MNRAS.311..576K}; \citealt{2005ApJ...625L..71H}; \citealt{2009ApJ...690...20S}).

Studying the impact of environmental factors on AGN fraction necessitates an investigation into the evolution of galaxies, clusters, and inter- and intracluster medium over cosmic time. Several studies show that dynamically evolving cluster environments over the history of the universe have affected AGN fractions in both clusters and fields (\citealt{2007ApJ...664L...9E}; \citealt{2017MNRAS.465.2531B}). In the local universe, there is evidence for anti-correlation between AGN fraction and galaxy density (e.g. \citealt{2017MNRAS.472..409L}), at least for luminous AGNs. Other studies have found comparable AGN fractions in clusters and fields for low-luminosity quasars (\citealt{2010ApJ...723.1447H}). However, at higher redshifts, the AGN evolution has been seen to follow a different evolutionary path (\citealt{2013ApJ...768....1M}). Cosmic conditions, such as galaxy and cluster morphologies, presence of denser ICM, dominance of dark matter etc. at higher redshifts have greatly impacted the current state of the universe and large-scale matter distribution.

We look at one of the largest cluster samples to study the relative fraction of AGNs in cluster galaxies and fields. The aim of this paper is to investigate the AGN fraction in clusters relative to their field regions in the local universe (redshifts less than 0.5) using X-ray, optical, and mid-IR surveys. This is crucial to understand how the different local and global environmental factors found in clusters and low density fields may affect the AGN activity (by triggering or suppressing AGNs). 

The paper is divided into the following sections: we describe the data and the photometric surveys used in Section \ref{sec:sec2}. In Section \ref{sec:sec3}, we describe the methodology used and in Section \ref{sec:sec4}, we discuss the results. Finally, in Sections 
\ref{sec:sec5} and \ref{sec:sec6}, we present the discussion and conclusions to our study respectively. We adopt the values for the cosmological parameters as follows: $\Omega_{m}$ = 0.3, $\Omega_{\Lambda}$ = 0.7, $H_0$ = 70 km$s^{-1}$$Mpc^{-1}$.

\section{THE DATA} \label{sec:sec2}
This study is based on 25801 galaxies in X-ray selected galaxy clusters and groups and 242972 field galaxies. This makes it one of the largest cluster catalogs used for studying the relative fraction of AGNs in high-density cluster environments against their surrounding low-density fields. In the following subsections, we describe the data.
\subsection{Cluster Sample}
The clusters studied in this paper are from the Meta-Catalog of X-ray selected galaxy clusters (MCXC), described in \citet{2011A&A...534A.109P}. It is partially based on the \emph{ROSAT} All-Sky Survey, abbreviated RASS hereafter. The majority of the RASS data was obtained by Position Sensitive Proporional Counter (PSPC) in the scanning mode with a count rate $\geq$ 0.06 counts/s. The catalog contains clusters from the surveys NORAS (Northern \emph{ROSAT} All-Sky galaxy cluster survey) (\citealt{2000ApJS..129..435B}) and REFLEX (\emph{ROSAT}-ESO Flux Limited X-ray Galaxy Cluster Survey) (\citealt{2004A&A...425..367B}). REFLEX surveys the southern hemisphere at declinations below +2.5\degr{} with a mean exposure of 335 s. It has a total coverage area of 13924 $deg^2$, which excludes the Galactic plane and Magellanic clouds. NORAS covers the northern sky above 0.0\degr{} declination and Galactic latitude $\geq$ 20\degr, with a mean exposure of 397 s. Additional data from the SGP survey covers a region of 1.013 steradian around the South Galactic Pole (\citealt{2002ApJS..140..239C}). The other part of the catalog is taken from serendipitous cluster catalogs, such as WARPS (\citealt{2002ApJS..140..265P}; \citealt{2008ApJS..176..374H}), EMSS (\citealt{1990ApJ...356L..35G}), 160SD (\citealt{2003ApJ...594..154M}), 400SD (\citealt{2007ApJS..172..561B}), and SHARC (\citealt{2000ApJS..126..209R}; \citealt{2003MNRAS.341.1093B}), that take observations of deeper pointed X-ray sources. There are five more contiguous RASS surveys and the five serendipitous surveys (mentioned above) used to create this catalog (\citealt{2011A&A...534A.109P}).

\begin{figure}
\includegraphics[width=8cm, height=8cm]{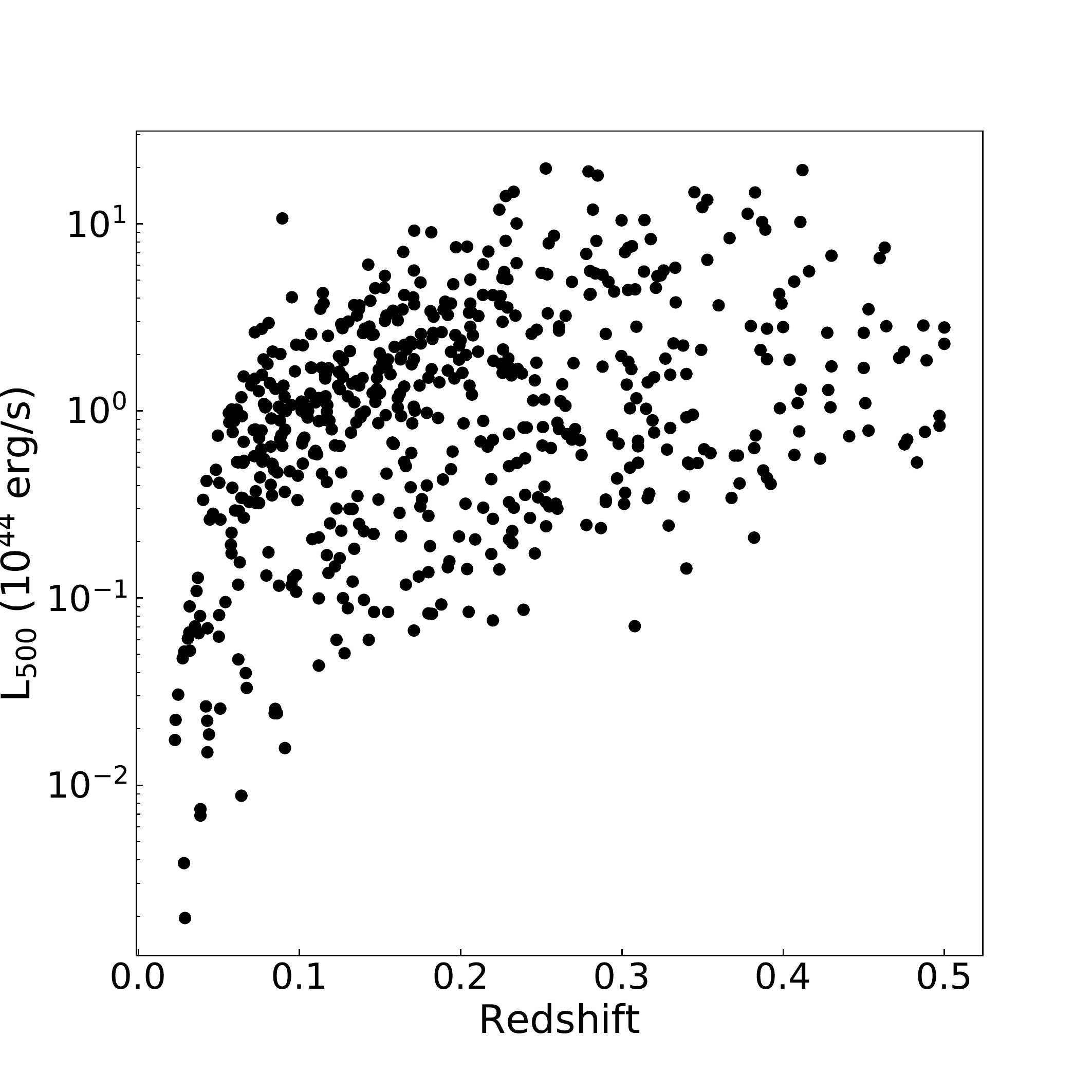}
\centering
\caption{Luminosity ($L_{500}$) versus redshift of the X-ray clusters used in this study. \label{fig:f6}}
\end{figure}

The Meta-catalog of X-ray selected galaxy clusters provides the cluster redshift, standardized 0.1 -- 2.4 keV band luminosity \lum, cluster mass \mass, and radius \rad, for 1743 clusters. The subscript 500 denotes the characteristic radius within which the mean density of the cluster is 500 times the critical density of the universe at the  cluster redshift. The majority of MCXC cluster are at low to medium redshifts, with a redshift peak at $z = 0.08$. About half of the clusters (49\%) have 0.1--2.4 keV band luminosities larger than \tenf\ erg\psec, with a mean luminosity, \lum\ = 2.11$\times$\tenf\ erg\psec. 
The cluster sample used in this study resembles a combination of multiple flux-limited samples where a large fraction of the clusters is derived from the RASS. 
Figure \ref{fig:f6} shows the luminosity distribution with respect to cluster redshift, where we can see the multiple flux limits. This distribution approximately contains clusters brighter than \lum\ = 10$^{43}$~erg\psec\ at all redshift values. There are a handful of low luminosity clusters below this luminosity value for $z<0.1$, and the same redshift bin has a deficit of the brightest clusters, at luminosities around \lum\ = 10$\times$\tenf\ erg\psec, because of the lack of survey volume for the lowest redshift bin. However, as we see in Section~\ref{sec:sec4}, our mass-based classification of clusters (mass and  luminosity are tightly correlated for clusters) does not show a significant difference in AGN fractions between the high- and low-mass categories in each redshift bin. Therefore, we argue that the flux-limited type of sampling from multiple surveys does not affect the AGN fraction results from this paper significantly.  

The mean mass, \mass\, for these clusters is 2.28$\times$10$^{14}$ \msun. The catalog contains clusters ranging from galaxy groups to rich clusters with the mass range between 0.96$\times$10$^{12}$ to 2.21$\times$10$^{15}$ \msun. Detailed methodology to calculate X-ray luminosities (\lum) and masses (\mass) for the galaxy clusters, along with their uncertainties, can be found in \citet{2011A&A...534A.109P}. We have selected 580 X-ray selected galaxy clusters for this analysis that lie in the redshift range 0.0 and 0.5 and are a part of the SDSS footprint. The average X-ray luminosity for our sample of clusters in  0.1--2.4 keV band is 1.77$\times$\tenf{} erg\psec and the average mass is 2.08$\times$10$^{14}$ \msun. These are moderately rich clusters with a large fraction being galaxy groups and thus not well studied in literature. 

\subsection{SDSS and WISE Galaxy Catalogs}
The galaxies were selected using the Sloan Digital Sky Survey data release 14 (SDSS DR14) (\citealt{2018ApJS..235...42A}). SDSS is a multi-band imaging and spectroscopic survey that uses a 2.5m telescope with a survey area of 14555 square degrees. It uses five broad bands ($u$,$g$,$r$,$i$, and $z$) where the average exposure time for each band is 53.9 s and the median redshift is 0.1 for the photometrically selected galaxies. This makes it a good catalog for the study of low redshift clusters. The median  5$\sigma$ depths for photometric images are 22.15, 23.13, 22.70, 22.20, and 20.71 mag for u, g, r, i, and z bands, respectively. 

WISE (Wide-field Infrared Survey Explorer) is a multi-band, mid-IR all-sky survey that was launched on 2009 December 14 and completed its first complete sky coverage on 2010 July 17 (\citealt{2010AJ....140.1868W}). WISE has imaged in four passbands 3.4, 4.6, 12, and 22$\micron$ (W1, W2, W3, and W4, respectively) with mean exposure times of 7.7 s (in W1 and W2 bands) and 8.8 s (in W3 and W4 bands). The 5$\sigma$ sensitivity in photometry is estimated to be 0.068, 0.098, 0.86, and 5.4 mJy at 3.4, 4.6, 12 and 22$\micron$, respectively. Saturation affects photometry for sources brighter than approximately 8.1, 6.7, 3.8, and $-$0.4 mag at 3.4, 4.6, 12, and 22$\micron$, respectively. 

\section{Methodology} \label{sec:sec3}
 Cluster regions were defined to encompass the entire angular extent from the cluster center out to \rad{}. The typical \rad{} value for our clusters at these redshifts is about 2 Mpc and the angular size ranges from 1\arcmin\ for the most distant clusters to 58\arcmin\ for the nearby clusters. We then selected SDSS galaxies within the cluster region and made a redshift cut using the photometric redshift values available for the SDSS galaxy data within $z_{cl}$ (1$\pm$0.05). For this study, we only select the galaxies in the absolute magnitude range $-20.0$ to $-22.0$. This luminosity range is detectable for redshifts considered in our analysis $z<0.5$ given the SDSS $r$ band limits of 22.2 mag.
 This range is chosen for all redshifts to avoid the selection bias toward low luminosity galaxies at low redshifts, and the range provides maximum number of galaxies for a statistical study of variation of AGN fraction with redshift.
 We bin the clusters with a 0.1 redshift bin, which is a factor of 2--3 of the mean uncertainties for photometric redshifts from 0.03 for $z < 0.3$ to 0.06 for $z < 0.5$ (Abolfathi et al.\ 2018).
 
 For each cluster, we uniquely defined a local field region, 5$\times$\rad $-$ 10$\times$\rad{} for comparison. Each SDSS galaxy in the dataset was identified uniquely with a single cluster or field region using SDSS photometric redshift (spectroscopic redshift was used where available) of the galaxy. This photometric redshift criterion will include foreground and background galaxies due to larger uncertainties in photometric $z$ values. Therefore, we also investigate relative AGN fraction for more robust analysis. The galaxies in SDSS DR14 were then matched with sources from the WISE All Sky Survey with a matching radius of 2\arcsec{} \citep[e.g.,][]{2015ApJS..218....8D} to obtain the infrared colors. AGN fraction, $f_A$ is defined as the ratio of the number of AGNs and total galaxies in the cluster.    

\begin{figure}
\plottwo{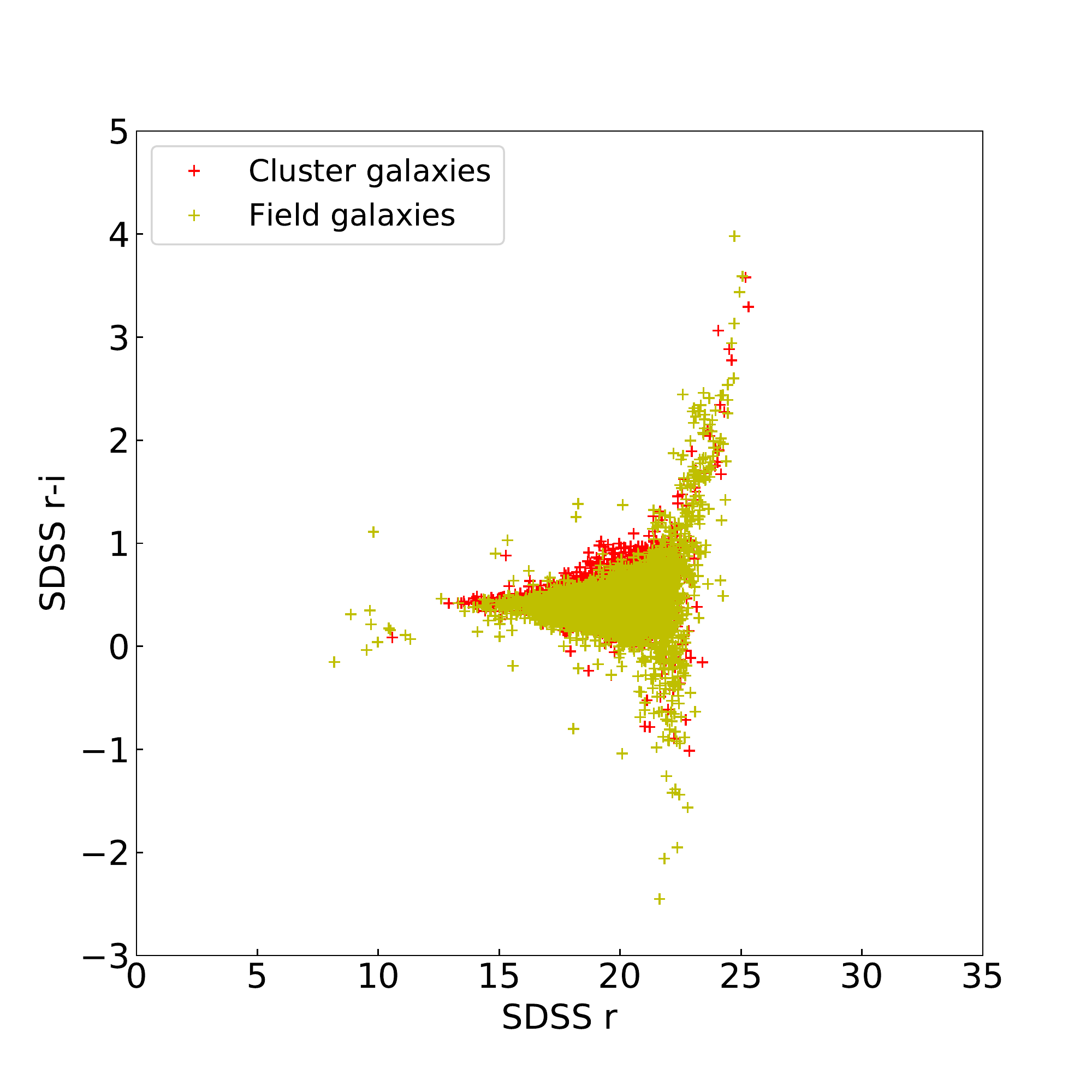}{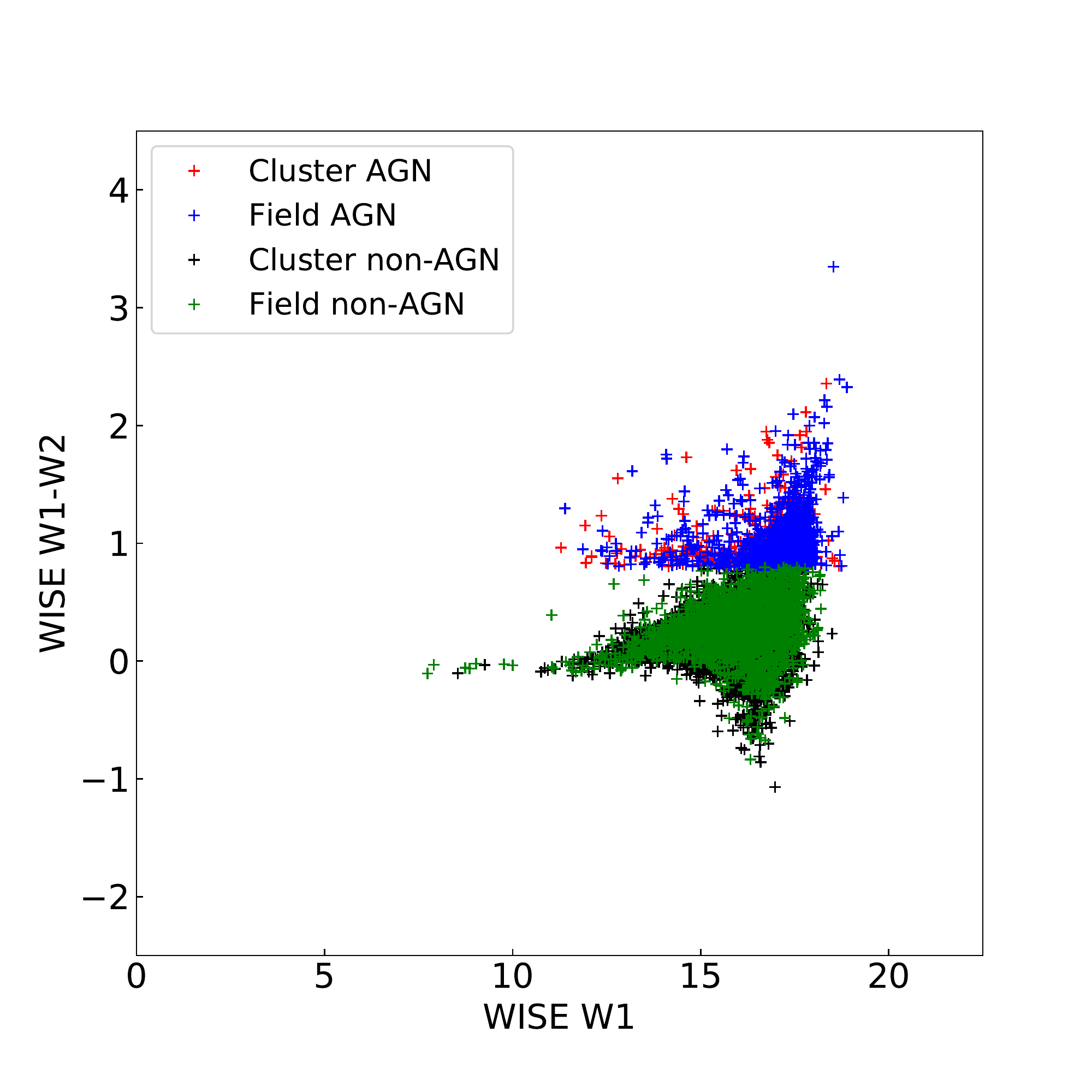}
\centering
\caption{Color-magnitude diagrams for our galaxy sample for redshift range 0.0--0.5. Left panel: Optical color-magnitude diagram in SDSS $r$ and $i$ bands for all cluster galaxies (red) and all field galaxies (yellow). Right panel: Mid-IR color-magnitude diagram using WISE $W1$ and $W2$ magnitudes for cluster AGNs (in red), field AGNs (in blue), cluster non-active galaxies (in black), and field non-active galaxies (in green). \label{fig:f1}}
\end{figure}

While several color cuts have been proposed in literature to select mid-IR AGNs, our AGN sample was selected using the mid-IR color selection criterion described in \citet{2012ApJ...753...30S}. Up to redshifts $\sim$ 3.5, mid-IR AGNs have been observed to have mid-IR colors between W1 and W2 filters (W1$-$W2) (\citealt{2012ApJ...753...30S}), with a different color distribution than galaxies, because they have a distinct characteristic in the spectral energy distribution (\citealt{2005ApJ...631..163S}). Extinction from dust can cause extreme reddening in high redshift AGNs for optical and UV surveys and while AGNs selected in mid-IR are relatively insensitive to absorption, they need to contribute significantly to the total mid-IR luminosity to distinguish them from their host galaxy. Thus, identifying AGNs based on mid-IR colors is a robust method for heavily obscured, high-luminosity AGNs with z $\leq$ 3.5. We impose a color cut of W1 -- W2 $>$ 0.8 which is generally accepted to identify AGN dominated colors well. The optical and IR color-magnitude diagrams for the galaxies across all redshift intervals are shown in Figure \ref{fig:f1}. The right panel shows the mid-IR color magnitude plot for cluster and field AGNs and non-active galaxies. In the left panel of Figure \ref{fig:f1}, the optical colors are shown in SDSS-r and SDSS-i bands for a fraction of cluster and field galaxies (AGNs and non-active galaxies). As can be seen, the majority of the cluster and field galaxies lie on the red sequence. The optical selection of cluster and field galaxies is biased toward red galaxies which have a more prominent 4000\AA break allowing for more accurate determination of photometric redshifts. However, since this selection bias is seen in both cluster and field galaxies, the relative comparison of AGN fractions between cluster and field is not significantly affected. Hence, we perform a comparative study of AGN fractions for cluster and field environments. 

\section{Results} \label{sec:sec4}
\subsection{Dependence on environment}
We separate the clusters in five redshift bins with a bin size of 0.1. We then calculate the AGN fractions in the clusters and fields in each redshift bin by adding all the data in the bin. In addition, we also split the cluster galaxies into two mass ranges and define those as low mass and high mass clusters. The mass cut was determined such that there were about the same number of cluster galaxies above and below the mass cut. The low mass cluster can be thought of as a group-like environment at that redshift, whereas the high mass cluster represents the more massive galaxy clusters. We define AGN fraction as the ratio of number of AGNs in a cluster (or field) to the total number of galaxies. The AGN fractions are listed in Table \ref{tab:table1}. The uncertainties associated with the AGN fractions are Poisson errors with 1$\sigma$ confidence limit. As discussed in section \ref{sec:sec3}, the color cut used to select AGNs is generally reliable in selecting luminous mid-IR AGNs at low redshifts. Although our absolute AGN fractions for clusters and fields are affected by magnitude dependent AGN selection function, the relative AGN fraction between clusters and fields is a more robust quantity because it largely cancels out the selection function. In the left panel of Figure \ref{fig:f2}, we see that AGN fraction is lower in cluster galaxies than that of field galaxies for each redshift bin. This result agrees with other studies in the literature that look at the environmental dependence of AGNs in clusters and fields (e.g., \citealt{2017MNRAS.472..409L}; \citealt{2018A&A...620A..20K}). We infer that at low redshifts, $z \leq 0.5$, relatively more AGNs are found in low density field environment surrounding galaxy clusters compared to the regions closer to the core of the clusters. 

The right panel of Figure \ref{fig:f2} shows further classification into low and high mass objects where clusters (high-mass objects) have log$_{10}$(\mass/\msun) $>$ 14.3 and groups (low mass objects) have log$_{10}$(\mass/\msun) $<$ 14.3. We choose the mass ranges to have comparable number of objects above and below the mass cut. AGN fractions for both mass ranges are very similar within the 1$\sigma$ uncertainties. The AGN fractions for group-like objects are slightly higher than cluster-like objects, except for $0.1 \leq z \leq 0.2$ bin. Based on the result seen in left panel, we expect low mass objects to have fractions somewhere between high mass clusters and the fields, which we modestly observe, limited by the signal-to-noise ratio of the data. The salient feature we observe is a significantly higher AGN fraction in field than either low or high mass clusters which is the same environmental dependence as seen in the left panel of Figure \ref{fig:f2}.

\begin{deluxetable}{lcccr}[b!]
\tablecaption{AGN fraction in clusters and fields for the five redshift bins \label{tab:table1}}
\tablecolumns{5}
\tablenum{1}
\tablewidth{0pt}
\tablehead{
\colhead{Redshift\tablenotemark{a}} & 
\colhead{Cluster} &
\colhead{Field} & 
\colhead{Low mass cluster} &
\colhead{High mass cluster} \\
}
\startdata
0.0 $\leq$ $z$ $\leq$ 0.1 & $0.0028_{-0.0009}^{+0.0013}$ & $0.0143_{-0.0009}^{+0.0009}$ & $0.0015_{-0.0006}^{+0.0010}$ & $0.0035_{-0.0015}^{+0.0024}$\\
0.1 $\leq$ $z$ $\leq$ 0.2 & $0.0058_{-0.0009}^{+0.0011}$  & $0.0190_{-0.0005}^{+0.0005}$ & $0.0084_{-0.0013}^{+0.0015}$ & $0.0068_{-0.0017}^{+0.0022}$\\
0.2 $\leq$ z $<$ 0.3 & $0.0286_{-0.0033}^{+0.0036}$ & $0.0745_{-0.0013}^{+0.0013}$ & $0.0263_{-0.0035}^{+0.0039}$ & $0.0425_{-0.0095}^{+0.0113}$\\
0.3 $\leq$ z $<$ 0.4 & $0.0399_{-0.0055}^{+0.0062}$ & $0.0791_{-0.0020}^{+0.0020}$ & $0.0390_{-0.0063}^{+0.0071}$ & $0.0501_{-0.0127}^{+0.0156}$ \\
0.4 $\leq$ z $<$ 0.5 & $0.0413_{-0.0090}^{+0.0108}$ & $0.0821_{-0.0033}^{+0.0034}$ & $0.0403_{-0.0103}^{+0.0127}$ & $0.0437_{-0.0176}^{+0.0250}$ \\
\enddata
\tablenotetext{a}{All the data for cluster and field in each $z$ bin is added }
\end{deluxetable}

\subsection{Dependence on redshift}
\begin{figure}
\plottwo{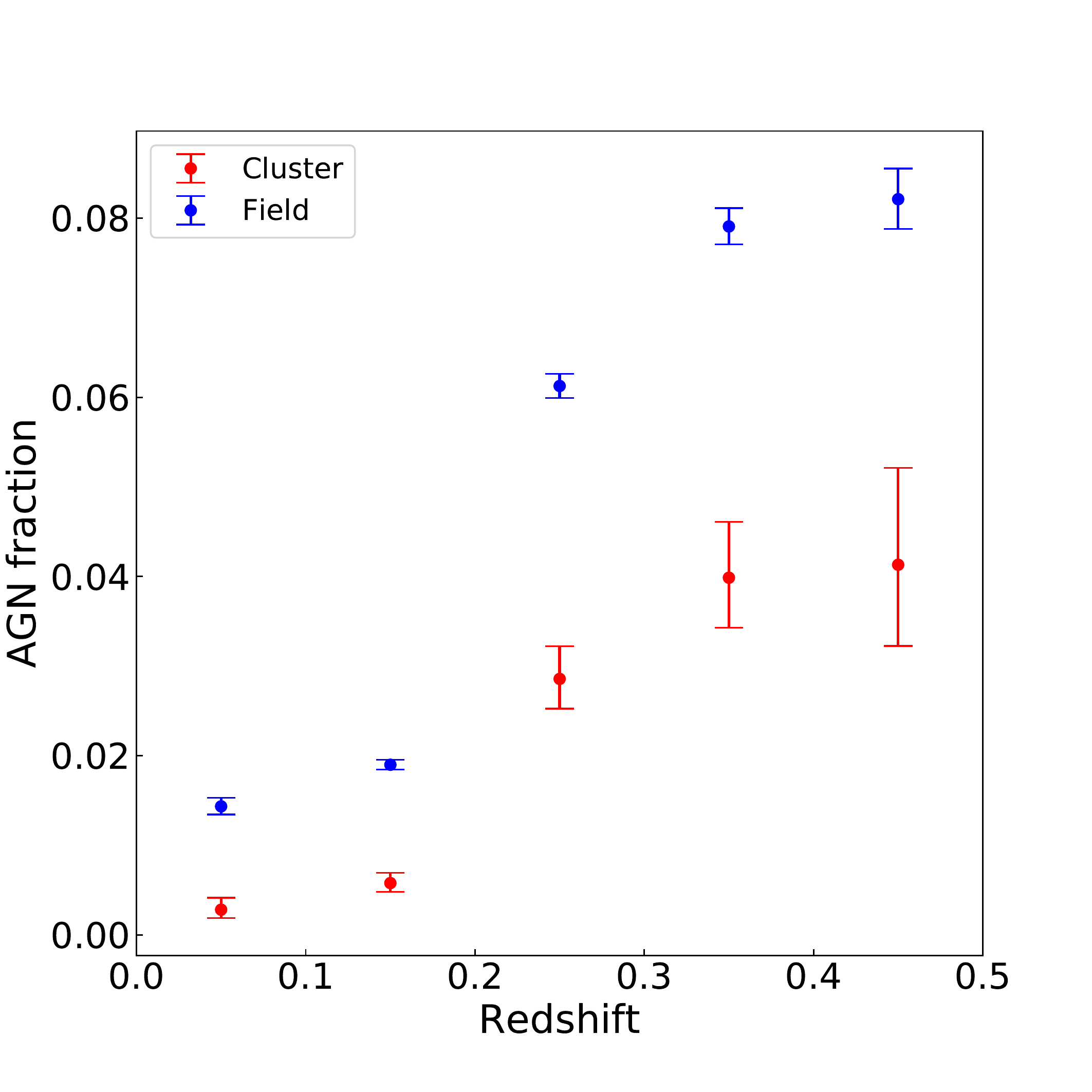}{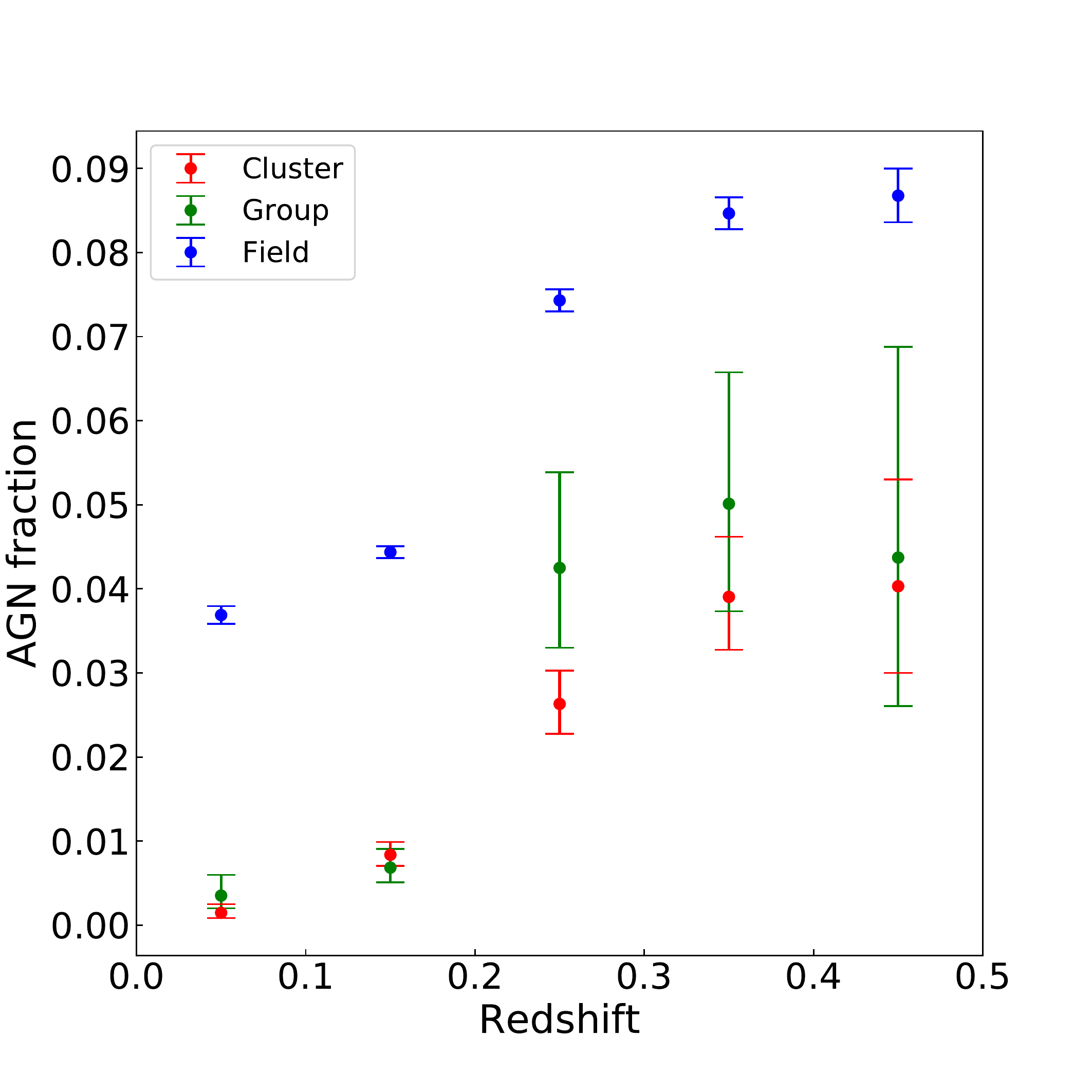}
\caption{Fraction of AGNs in high and low density environments as a function of redshift. Left panel: AGN fraction for redshift bins from $z$=0.0 to $z$=0.5, with bin size=0.1 for cluster and field objects. The errors are Poisson errors that correspond to 1$\sigma$ confidence limits. Right panel: Same as the left panel, but we divide cluster galaxies into low mass objects (green circles) and high mass objects (red circles). The blue circles are field galaxies in both panels.\label{fig:f2}}
\end{figure}

 As described in Section \ref{sec:sec3}, we divided the clusters into various redshift bins to study the fraction of AGNs as a function of redshift. First, we compared AGN fractions in cluster and field for each $z$ interval, which is plotted in Figure \ref{fig:f2}, left panel. For our relatively low redshift range, it can be seen that the mean cluster AGN fraction increases as we go from $z < 0.1$ to $z < 0.5$, however, the value flattens out for $z > 0.4$ within 1$\sigma$ errors. A similar trend of AGN fraction with $z$ is seen for field galaxies as well. The comparable $f_A$ values within error bars for field galaxies for $z<0.4$ and $z<0.5$ could be a result of the significantly smaller samples of both cluster and field galaxies at higher redshifts. Similar trend has been found in previous studies, such as \citet{2013ApJ...768....1M} who studied AGN fraction at high redshift ($z$ $\sim$ 1--1.5) and found an increase in X-ray AGN fraction from $z=0$ to $z=3.0$ for cluster and field AGNs. There is a 14.7 times increase in cluster AGN fraction from $z<0.1$ to $z<0.5$. Similarly, for field region, AGN fraction increases 5.7 times from $z<0.1$ to $z<0.5$. The percentage increase in cluster AGN fraction is higher than that in field AGN fraction.   
The right panel of Figure \ref{fig:f2} shows the evolution of low and high mass cluster AGN fractions with redshift. For high mass clusters, there is a positive trend with $z$ up to $z=0.3$ and then the values flatten out. For low mass objects, the dependence between AGN fraction and redshift follows a similar trend as high mass objects. However, the error bars are bigger for low mass objects. For field galaxies, the plot is the same as the left panel. In general, we observe an increasing AGN fraction with redshift as found in many studies done previously (e.g. \citealt{2005A&A...441..417H}; \citealt{2013ApJ...768....1M}).  
 
\begin{figure}
\includegraphics[width=8cm, height=8cm]{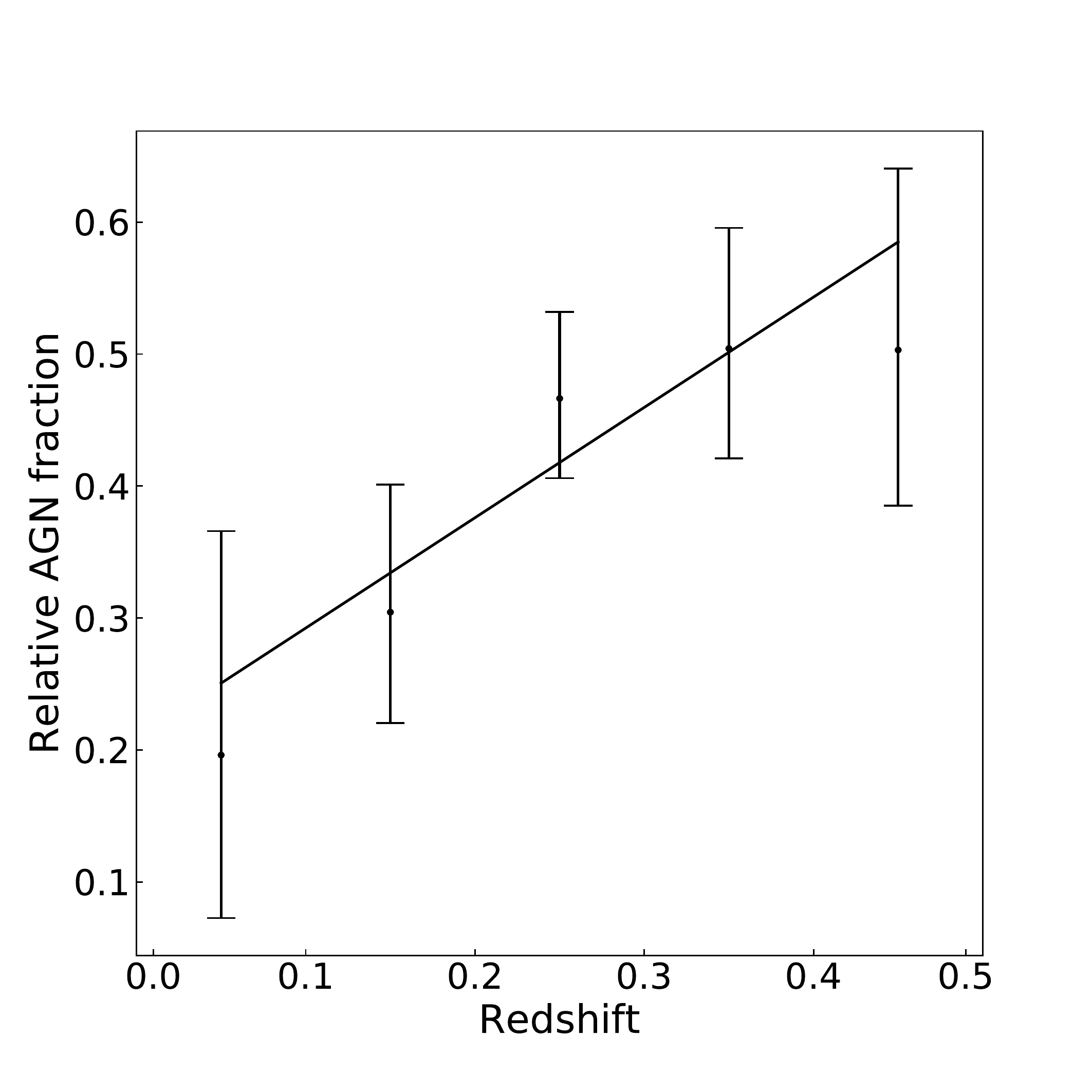}
\centering
\caption{Relative AGN fraction for cluster and field galaxies as a function of redshift. The errors are Poisson errors that correspond to 1$\sigma$ confidence limits. The best-fit line has a slope of $0.84\pm0.24$. \label{fig:f3}}
\end{figure}

Figure \ref{fig:f3} shows relative AGN fraction versus redshift. Relative fraction is more robust against selection effects, because they largely cancel out between field and cluster fractions. The correlation is determined using a best fit line and a positive slope of $0.84\pm0.24$, a 3.5$\sigma$ deviation from zero slope, which indicates that the relative AGN fraction also increases with redshift as is the case of absolute AGN fractions at $z<0.5$. 
  
\subsection{Dependence on angular cluster-centric distance}
\begin{figure}
\includegraphics[width=10cm, height=20cm]{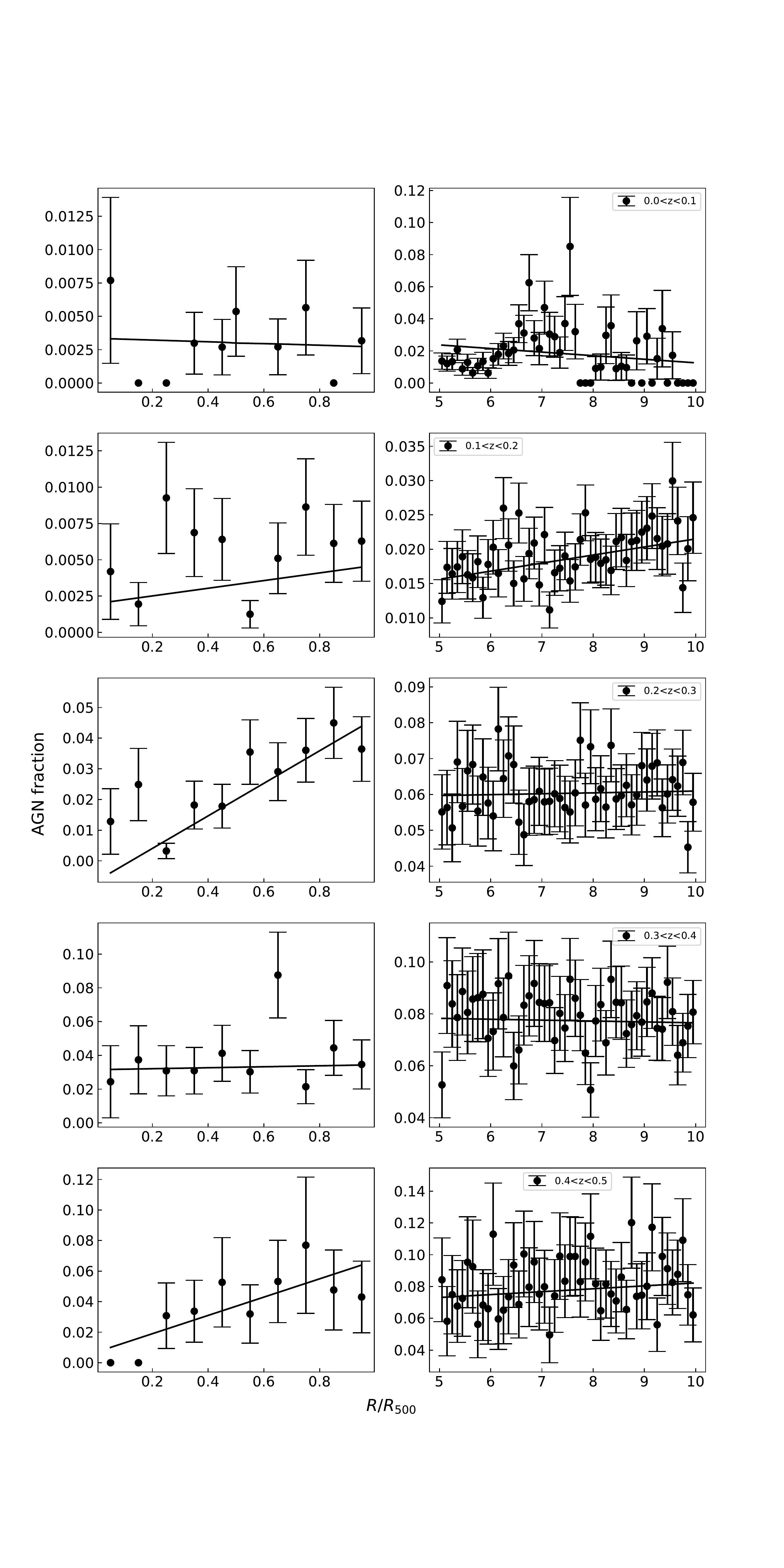}
\centering
\caption{Fraction of AGNs as a function of angular clustercentric distance (in terms of \rad{}). Left panel: $f_A$ for cluster galaxies for the five $z$ bins. Right panel: Same as the left panel, but for field galaxies. The errors are Poisson errors that correspond to 1$\sigma$ confidence limits. \label{fig:f4}}
\end{figure}

We also investigated the dependence of AGN fraction on normalized radial offset from the center of the cluster for cluster and field galaxies. The distances are angular distances in \rad{}. In the left panel of Figure \ref{fig:f4}, AGN fraction is plotted as a function of distance for cluster galaxies for the five redshift intervals. We see an increase in $f_A$ from the core of the cluster out to \rad{} for all the intervals. The slopes and $1 \sigma$ uncertainties for the best-fit lines are $0.016 \pm 0.004$, $0.003\pm 0.003$, $0.053 \pm 0.011$, $0.003 \pm 0.016$, and $0.060 \pm 0.021$ respectively. The dependence does not show a trend with increasing redshift. We see the slope decrease at first, then increase, then decrease again for $0.3 < z < 0.4$. The trend line is the steepest for the highest redshift bin. This suggests the cluster evolution may not be significant from $z = 0.5$ clusters to present day clusters to observe an evolution in cluster-centric dependence of AGN fraction.  

AGN fractions as a function of distance for field galaxies are in the right panel of Figure \ref{fig:f4}. As stated above, we define field galaxies as lying between 5$\times$\rad{} to 10$\times$\rad{}. There data is very scattered for field galaxies and the slopes are very close to 0 which indicates there is not a significant dependence on distance. We expect that the field fraction would not depend on the distance from the center of the cluster.

\subsection{Dependence on absolute magnitudes}
\begin{figure}
\includegraphics[width=10cm, height=20cm]{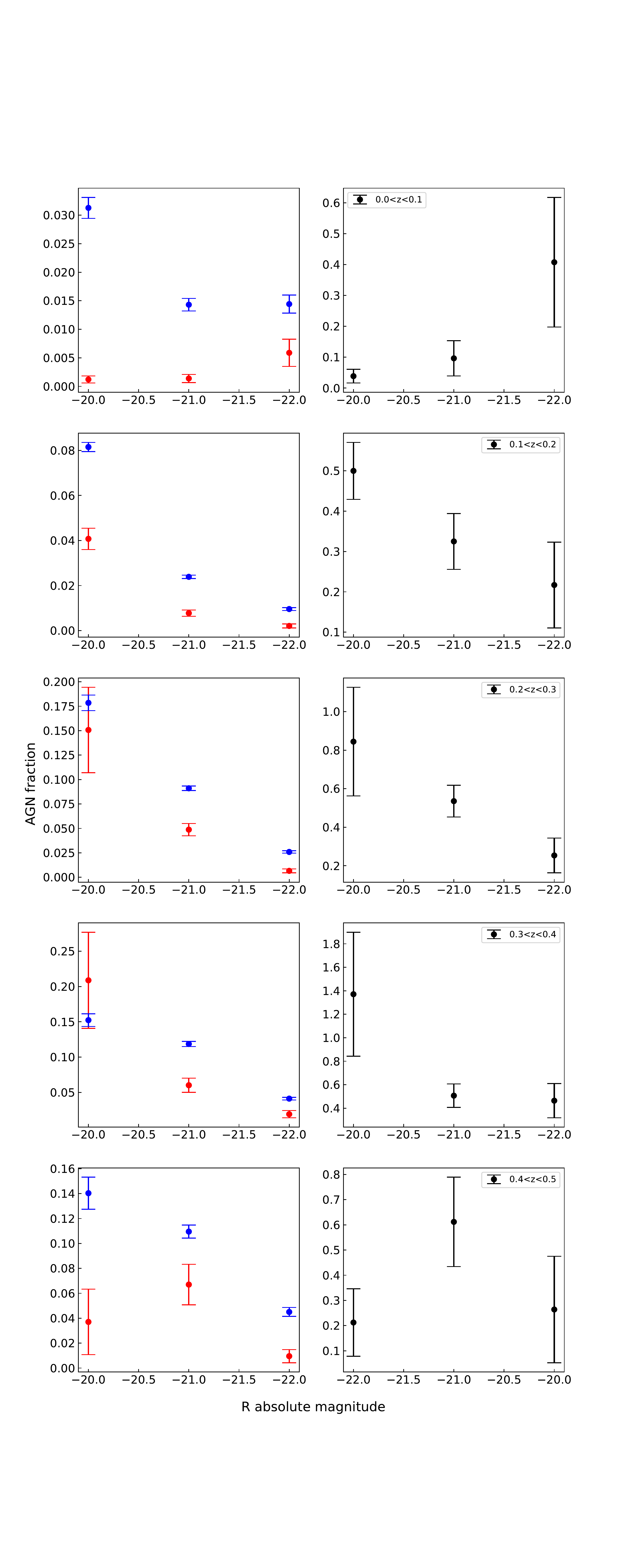}
\centering
\caption{Fraction of AGNs as a function of R absolute magnitude. Left panel: $f_A$ for cluster (red) and field (blue) galaxies for the five $z$ bins. Right panel: Same as the left panel, but for relative AGN fractions. The errors are Poisson errors that correspond to 1$\sigma$ confidence limits.\label{fig:f5}}
\end{figure}

We also investigate the dependence of AGN fraction on the \Mr{}, absolute magnitude in the SDSS-r band. The absolute magnitudes for SDSS r-band were calculated using the distance modulus equation, with the K-correction terms for specific redshift values extracted from \citet{2010ApJ...713..970A} using the spectral energy distribution template for elliptical galaxies in clusters. The template is in the wavelength range from 0.03 $\mu$m to 30 $\mu$m and is based on the NOAO Deep-Wide Field Survey Bootes which has multi-wavelength photometric observations and also the AGN and Galaxy Evolution Survey which provides the spectroscopic data. This template is extended into the mid-IR (\citealt{2008ApJ...676..286A}) and is useful as we select the AGNs based on the mid-IR color selection criteria mentioned in section \ref{sec:sec3}. Our luminosity cut represents intermediate-luminosity AGNs ($-22 <$ \Mr{} $< -20.0$). This luminosity cut was made to enable comparison of AGN fractions for lowest and highest redshift bins since the peak R-band luminosity shifts to lower absolute magnitudes for higher redshifts as only the most luminous AGNs would get selected at higher redshifts. Thus, analyzing $f_A$ for a wider magnitude range would result in a biased sample of fewer faint magnitude, high $z$ AGNs. Our magnitude range selects the majority of the AGNs in each redshift interval in our sample to avoid selection bias and therefore provides a more complete sample of cluster and field AGNs to calculate the AGN fraction. Because the selection functions cancel out between cluster and field AGN fractions, relative AGN fraction is a more robust measurement. Figure \ref{fig:f5} shows AGN fraction as a function of R absolute magnitude for all the redshift bins. In the left panel, we have shown AGN fraction dependence on absolute magnitude for cluster and field galaxies. While a general trend of decreasing fraction is seen for field galaxies for $0.1 < z < 0.5$, it is not significant. For $z < 0.1$, the AGN fraction decreases and then becomes constant. The correlation for cluster galaxies is not conclusive. While AGN fraction seems to increase for the lowest $z$ bin, it shows the opposite dependence for the next three bins. For $z < 0.5$, $f_A$ does not have a significant trend with absolute magnitude.

The right panel of Figure \ref{fig:f5} shows the same relationship for relative AGN fraction. No obvious trend is seen with absolute SDSS r-band magnitude. Similar to the left panel, the relative AGN fraction increases with magnitude for $0 < z < 0.1$. It then decreases with absolute magnitude for next two $z$ bins and flattens out for the highest redshift clusters in our sample. We expect a higher AGN fraction with increasing luminosity if we consider the absolute magnitude to be representative of cluster stellar mass. However, mid-IR selection criteria used in this study leads to a biased selection of AGNs as described in \citet{2012ApJ...753...30S} which selects WISE AGN candidates that peak around SDSS r $\approx$ 20.0. Very few AGN candidates selected are brighter than $\sim$ 19th mag in SDSS r-band (\citealt{2012ApJ...753...30S}). This selection bias would explain the scattered distribution of AGN fraction relative to R absolute magnitude.

\section{Discussion} \label{sec:sec5}
We discuss some of the selection biases and the robustness of our results. For the mid-IR color selection of AGNs, we report the results for the selection we use, W1-W2 $> 0.8$. However, we also selected different color cuts above and below W1-W2 $= 0.8$ and found that while absolute AGN fraction values changed, the trends seen with various cluster and galaxy properties and relative fractions are robust against the color selection criterion. The field fractions are higher than cluster fractions at all redshifts. The optical selection of cluster and field galaxies is biased toward red galaxies as described in section \ref{sec:sec3}. However, comparing cluster AGN fraction against field fraction makes this bias unimportant as we are selecting more red galaxies in both cluster and field. 

As discussed in section \ref{sec:sec4}, our results show that AGN fraction in clusters is lower than that in fields at all redshifts considered in this study. This is in agreement with several AGN fraction studies in massive clusters done previously (e.g. \citealt{2004MNRAS.353..713K}; \citealt{2017MNRAS.472..409L}). This indicates cold gas inflow toward the center where the black hole is located is more efficient for isolated galaxies situated in fields, via secular processes and/or mergers. The role of mergers in being the dominant triggering mechanism for AGN activity has been debated for a long time. If low-z AGNs are turned on predominantly from the inflow of gas during major galaxy interaction and merger events (e.g. \citealt{1988ApJ...325...74S}), it would be expected that high velocity dispersions in moderate to rich clusters would decrease the efficiency of these mergers resulting in a lower fraction of active galactic nuclei found in inner regions of clusters. Semi-analytical modeling of dynamically young, evolving galaxy groups formed by mergers of smaller galaxy systems is shown to have higher black hole accretion rate than old, relaxed galaxy groups formed from major mergers at higher $z$ (\citealt{2014MNRAS.442.1578R}; \citealt{2019MNRAS.486.1509R}), leading to more efficient AGN activity and thus, higher AGN fractions. This points toward a reduced merger efficiency in the richest galaxy conglomerations. Outer regions of clusters and low density fields in the immediate vicinity of the clusters are seen to have higher fraction of AGNs at low redshifts, attributed to more efficient mergers while AGN activity is strongly suppressed in the inner regions of rich clusters (e.g. \citealt{2010ApJ...714L.181K}; \citealt{2018A&A...620A..20K}). We find a significant difference between cluster and field AGN fractions for both the cases, when we divide our clusters based on mass and when we do not, which is clearly indicative of an environmental dependence for these intermediate-mass clusters. Our conclusions are also supported by \citet{2008ApJ...673..715C} who found an anti-correlation between AGN activity and local density for moderately bright galaxies (\Mr{} $\approx -20$) similar to the galaxies used in our study. Many theoretical studies that model galaxy interactions have found a close association between galaxy mergers and accretion of cold gas onto the supermassive black hole, thus triggering AGN activity (e.g. \citealt{2003ApJ...593...56D}; \citealt{2005Natur.433..604D}; \citealt{2006MNRAS.367..864C}). Role of mergers and hence the environment of these AGNs is also very heavily dependent on AGN selection criteria (\citealt{2016A&A...592A..11K}; \citealt{2018PASJ...70S..37G}), luminosity of the galaxy sample (\citealt{2018A&A...620A...6M}; \citealt{2019PASJ...71...31C}), and redshift (\citealt{2009ApJ...701...66M}; \citealt{2018PASJ...70S..30S}). Many studies that show no correlation between AGN activity and environment are done at high z and typically use X-ray AGNs (e.g. \citealt{2009ApJ...695..171S}; \citealt{2017geat.confE...9V}). Our results agree with the recent study done by \citet{2019MNRAS.487.2491E} which looks at the role of interactions in triggering mid-IR selected AGNs in the local universe by studying morphological disturbances in these galaxies. They find that in 59\% of mid-IR AGNs, the morphology indicates interaction undergone by the galaxy and conclude that mergers seem to play a dominant role in fueling the active galactic nuclei. In their study, \citet{2019MNRAS.487.2491E} find different results for merger-driven triggering for optical AGNs versus mid-IR selected AGNs indicating selection bias in looking at the environment dependence on the triggering of active galaxies between various studies done in the literature. Ram pressure stripping would also play an important role in stripping galaxies of their gas, especially close to the cores of these clusters (\citealt{1972ApJ...176....1G}; \citealt{2014ApJ...781L..40E}). Pressure from the hot gas in the intracluster medium drives the gas out of the galaxy, including the denser cold gas which feeds the central engine.

Our result for group and cluster classification based on \mass{} shows comparable AGN fractions for the two mass classes. We expect low mass galaxy groups to have an environment and properties, such as stellar mass density similar to the outskirts of massive clusters. For the local universe, we expect groups to have higher AGN fraction than rich galaxy clusters due to an enhancement in triggering opportunities, especially because of lower relative velocities of galaxies in groups. We find slightly higher values for group-like objects, however, within the error bars, the difference is not significant (see Figure \ref{fig:f3}, left panel). Most clusters selected from the MCXC catalog are intermediate-mass to rich clusters. Therefore, the low-mass groups we have selected (log$_{10}$(\mass/\msun) $<$ 14.3) are still relatively massive objects and we conclude that our calculated AGN fraction for the low-mass category may not be truly representative of small galaxy groups. For example, \citet{2017MNRAS.472..409L} find significantly higher AGN fraction for high mass systems than low mass systems. However, it should be noted that the mass cut they use is log$_{10}$(M$_*$/\msun) $>$ 10.6, which is much lower than our cluster mass range. 
Many studies support the broad idea that the population of active galaxies found in galaxy clusters and groups and low tidal density field regions reduces from early universe ($z \sim 1-2$) to present day. AGN fraction evolution with redshift shows a positive dependence as seen in Figures \ref{fig:f2} and \ref{fig:f3}. Our results support this theory as we see an increase in mean cluster and field $f_A$ from $z=0.1$ to $z=0.5$. We infer our comparable $f_A$ values for $z=0.4$ and $z=0.5$ within error bars are a result of larger uncertainties in photometric redshifts and a smaller sample of high $z$ clusters and member galaxies. Several factors, such as reduced efficiency of massive mergers due to high relative velocities of member galaxies in clusters (\citealt{1997ApJ...481...83M}; \citealt{2007MNRAS.375.1017A}), smaller fraction of gas-rich galaxies (\citealt{2013ApJ...768....1M}), and ram pressure stripping (see \citealt{2017Natur.548..304P}; \citealt{2018MNRAS.474.3615M}), affect the availability of cold gas that fuels the accreting black hole in the local universe. Hydrodynamical and N-body simulations suggest an increase in merger rate with redshift for field galaxies (\citealt{2015MNRAS.449...49R}) which agrees with our result of higher AGN fraction in field at higher redshifts. At high redshifts, protoclusters provide a dense galaxy-rich environment which increases the probability of mergers in galaxies rich in gas aiding in fueling the central black holes and turning on the nuclear activity. This indicates a scenario for changing mechanisms for nuclear activity triggering over cosmic time where there is a transition from high-luminosity mergers dominant out to $z \sim 1.5$ to more secular processes, such as minor mergers, tidal effects, playing a more significant role in the local universe (\citealt{2009ApJ...701...66M}). Our findings for enhanced cluster (high-mass) and group (low-mass) AGN fractions at higher $z$ values is consistent with previous studies done in the literature that have found mergers of luminous, red galaxies in galaxy clusters and groups up to $z \sim 1$ and lower merger fractions in low redshift clusters (e.g. \citealt{1998ApJ...497..188C}; \citealt{2005ApJ...627L..25T}). By present day, cluster galaxies are believed to have been depleted of a significant amount of their cold gas relative to field galaxies, resulting in a smaller relative fraction of these active galaxies. Higher number of AGNs relative to total galaxies have been found at high redshifts (e.g. \citealt{2007ApJ...664L...9E}; \citealt{2010ApJ...723.1447H}; \citealt{2013ApJ...768....1M}). 

Studies have also found AGN fractions to vary with clustercentric distance (\citealt{2014MNRAS.437.1942E}; \citealt{2019A&A...623L..10K}). Our result for AGN fraction dependence on clustercentric distance shows a higher rate of AGN occurrence from as we go from cluster core to the outer regions of the cluster ($R \sim$ \rad{}) for all the redshift bins between $0.0-0.5$. This increase in AGN fraction suggests the presence of different physical conditions in the cluster core and outskirts and might be indicative of more efficient AGN quenching processes, such as ram pressure stripping, and fewer mergers near the cluster potential where majority of the red and dead ellipticals with high velocity dispersions reside. This result is in agreement with the increasing density of X-ray AGNs from $0 <$ R/\rad{} $<$ 1 found by \citet{2019A&A...623L..10K} in massive (\mass{} $> 10^{14}\msun$), high redshift ($z \sim 1$) clusters. However, variation of AGN density with distance from the cluster center and in the outskirts has also been found to depend on mass range of the sample. For example, \citet{2018MNRAS.475.4223G} look at the distribution of AGN in the infall region of the cluster projected phase space compared to cluster core and find that in high-mass groups (log$_{10}$(M$_{200}$/\msun) $>$ 13.5), AGN prefer the infalling galaxy population whereas for low-mass groups, they do not find any difference in AGN fraction between core and infall region. To better understand the environment dependence of AGN fraction in galaxy groups versus large clusters, we will need a sample of low-mass galaxy groups to perform the same analysis. The lack of redshift evolution of this trend might indicate that the cluster evolution from $z = 0.5$ to $z = 0$ does not affect the rate of AGNs inside the cluster. However, it should be noted that the larger errors in galaxy members' redshifts may result in a bias at high $z$.

We obtain inconclusive results for the relation between AGN fraction and SDSS r-band absolute magnitudes for cluster galaxies. The field AGN fraction decreases with increasing absolute R magnitude in SDSS bands.
For $0.0 < z < 0.1$, the AGN fraction peaks at \Mr{} $\sim -20.0$ for field galaxies and levels out for galaxies brighter than \Mr{} $\sim -21.0$. Whereas for the remaining four $z$ intervals, $0.1 < z < 0.5$, there is a steady decline in AGN fraction from \Mr{} = $-20.0$ to \Mr{} = $-22.0$. This contradicts the result found in \citet{2010ApJ...723.1447H} where they compared field X-ray AGN fractions in absolute i-band magnitude bins and found higher AGN fraction for $-22 < M_i < -21$ than for $-21 < M_i < -20$. This might be due to selection bias in our mid-IR selected AGNs in the field where very few of the WISE-selected AGNs are brighter than SDSS-i $\sim$ 19. We only select a small range of absolute magnitudes to study intermediate luminosity AGNs to keep it consistent over all the redshift bins and find no significant trend for cluster active galaxies or the relative AGN fraction. Larger sample of AGNs over a wider luminosity range would be needed to understand if luminosity dependence of AGN fraction is a function of luminosity class.

\section{Conclusions} \label{sec:sec6}
We have presented a study of the fraction of active galaxies in clusters and fields in an attempt to understand the still-debated environmental dependence of nuclear activity in AGNs. We have calculated the fraction of AGNs in clusters and fields in five redshift intervals between redshifts 0.0 to 0.5. For each redshift bin, we find that the field AGN fraction is significantly higher than the cluster AGN fraction. This result is in agreement with several studies done in the literature where low density (field) regions were found to have higher AGN fraction than their overdense counterparts at low to intermediate redshifts. It suggests that fields provide an ideal environment for facilitating gas inflows during mergers because of the relatively lower velocity dispersions in galaxies, triggering nuclear activity in AGNs more efficiently than the dense clusters. As we go from low to high redshifts, the relative fraction of AGNs seems to flatten out. AGN fraction has previously been found to increase with redshift which indicates a significant depletion in the availability of cold gas in clusters and fields in present universe compared to the scenario at high redshifts ($z \sim 1-1.5$). The comparable AGN fraction values for $z=0.4$ and $z=0.5$ could be because of higher uncertainty due to much smaller sample of galaxies compared to lower redshift galaxies. We also find a strong dependence of cluster AGN fraction on clustercentric distance which might be indicative of the differences in galaxy evolution due to different physical conditions present in the cluster core and outskirts, futher reinforcing the environmental dependence of AGN activities.  
The field AGN fraction is almost constant with distance from the cluster center, showing that we have selected field regions sufficiently away from the clusters. The dependence of AGN fraction on $R$ absolute magnitude is not conclusive and we attribute mid-IR selection bias to be a reason due to the selection of relatively brighter AGN candidates. 
In future, we plan to increase our sample size with spectroscopic data and low-mass X-ray selected clusters to improve the statistical significance of our results to better understand the environmental dependence of AGN fraction.

\acknowledgements
We are grateful to the anonymous referee for the helpful comments and recommendations that made the paper clearer and stronger. 
We acknowledge the financial support from the NASA ADAP programs NNX15AF04G, NNX17AF26G, NSF grant AST-1413056. 
\end{document}